\documentclass[apj]{emulateapj}

\usepackage{apjfonts}

\usepackage{graphicx}

\def\apjl{ApJL}
\def\prd{PRD}
\def\prl{PRL}
\def\pr{PR}
\def\rmp{Rev.\ Mod.\ Phys.}
\def\apj{ApJ}
\def\aap{A\&A}
\def\nat{Nature}
\def\araa{Ann.\ Rev.\ A\&A}
\def\mnras{MNRAS}
\def\apjs{ApJ Suppl.}
\def\app{Astrop.\ Phys.}
\def\azh{Astronom.\ Zh.}
\def\aaps{Astrop. \& Space Science}
\def\grl{Geoph.\ Res.\ Let.}
\def\jgr{J.\ Geophys.\ Res.}
\def\pasj{Publ.\ of the ASJ}
\def\ZfNat{Z.\ f.\ Nat.}
\def\ZfA{Z.\ f.\ A.}
\def\ARNSci{ARN Science}
\def\ZNat{Z.\ Nat.}
\def\jcap{J. Cosmology Astropart. Phys.}
\begin{document}
\title{Cosmic ray transport and anisotropies} \shorttitle{Cosmic ray transport and anisotropies}
\author{Peter L.~Biermann\altaffilmark{1,2,3,4,5}, Julia K.~Becker\altaffilmark{6},
 \\Eun-Suk Seo\altaffilmark{7},
 Matthias Mandelartz\altaffilmark{6}}
\shortauthors{P.L.\ Biermann et al.}
\altaffiltext{1}{MPI for Radioastronomy, Bonn, Germany}
\altaffiltext{2}{Dept. of Phys. \& Astr., Univ. of Alabama, Tuscaloosa, AL, USA}
\altaffiltext{3}{Inst. Exp. Nucl. Phys., Karlsruher Institut f{\"u}r Technologie KIT, Germany}
\altaffiltext{4}{Dept. of Phys., Univ. of Alabama at Huntsville, AL, USA}
\altaffiltext{5}{Dept. of Phys. \& Astron., Univ. of Bonn, Germany}
\altaffiltext{6}{Ruhr-Universit\"at Bochum, Fakult\"at f\"ur Physik \&
 Astronomie, Theoretische Physik IV, D-44780 Bochum, Germany}
\altaffiltext{7}{Dept. of Physics, Univ. of Maryland, College Park, MD, USA}
\begin{abstract}
We show that the large-scale cosmic ray anisotropy at $\sim 10$ TeV can be explained by a modified Compton-Getting effect in the magnetized flow field of old supernova remnants. This approach suggests an optimum energy scale for detecting the anisotropy. Two key assumptions are that propagation is based on turbulence following a Kolmogorov law and that cosmic ray interactions are dominated by transport through stellar winds of the exploding stars. A prediction is that the amplitude is smaller at lower energies due to incomplete sampling of the velocity field and also smaller at larger energies due to smearing. 
 \end{abstract}
\parindent=0cm
\parskip=0.2cm

\maketitle

\section{Introduction}

In the past few years, a number of cosmic ray (CR) experiments have convincingly shown that the arrival directions of Galactic CR particles are not fully isotropic in the sky:
A large-scale anisotropy with a dipole and quadrupole component was first observed in the northern hemisphere at an energy of several TeV \citep{amenomori2006,guillian2007,abdo2008,abdo2009}. The same structure was later found in the southern hemisphere at a mean energy of 20 TeV \citep{abbasi2010}. The large-scale anisotropy $I$, observed at several to tens of TeV energies, is of a level $ I \sim 10^{-4}\; \mathrm{to} \; 10^{-3}$. The results are summarized in Table \ref{table1}.\\
\begin{table}[ht]%
\begin{center}%
\caption{The observed large-scale CR anisotropies. References are: [1] - \citet{abbasi2010}; [2] - \citet{abbasi2011b}; [3] - \citet{amenomori2006}; [4] - \citet{abdo2008}; [5] - \citet{abdo2009}.}
\begin{tabular}{r|ccccc}%
\hline%
\hline%
& \multicolumn{2}{c}{IceCube}& Tibet & Milagro & SuperK\\%
\hline%
Refs& [1]&[2]& [3]& [4,5]& [3]\\%
mean $E$ [TeV]& 20 & 400 & 4 - 12& 6 & 10\\%
$I$ [$10^{-4}$]&  8 & 4 & 4 & 4 & 7 \\%
hemisphere& S & S  & N & N & N \\%
phase [R.A. (deg)]&  50 & 240 & 65 & 190 & 35\\%
\hline%
\hline%
\end{tabular}%
\label{table1}%
\end{center}%
\end{table}%
Apart from the large-scale anisotropy, several experiments showed that at the same energy scale smaller angular scale excesses and deficits do exist, with extensions from a few degrees in the sky up to about 20 degrees in one direction \citep{abdo2009,abbasi2011a}. Most recently, it was shown that the large-scale anisotropy which is present at TeV energies vanishes in the southern hemisphere at around 400 TeV and a new component emerges instead, a clear deficit of a 20 degree scale and with an intensity of the order of $\lesssim \, 10^{-4}$ \citep{abbasi2011a}. While the deficit has a significance of $6.3\,\sigma$, a clear excess is not yet distinguishable from current statistics \citep{abbasi2011b}. What is important about this result is that it is clearly not a simple dipole component, which would be just symmetric in excess and deficit, but the anisotropies clearly display disjunct excesses and deficits.\\
First we may have to ask, why we should expect isotropy of cosmic rays at all: the essential answer was given by \citet{schlueter_biermann1950}: Magnetic fields get strengthened until they scatter cosmic rays into near prefect isotropy \citep[also found by][]{hanasz2004,hanasz2009}. On this basis they estimated the strength of magnetic fields to be of order 4~$\mu$G, an estimate, which has held up remarkably well \citep[e.g.][]{beck1996}. Thus we can expect isotropy in the reference frame of the local interstellar medium, which is very well coupled to the magnetic fields \citep[e.g.][]{appenzeller1974}, a connection which is kept by instabilities \citep{parker1966}.
However, the Sun was only coupled to the interstellar medium at birth, 4.5 billion years ago: Interaction with massive interstellar gas clouds slowly increases the peculiar velocity of stars with age \citep[e.g.][]{julian1967,wielen1975}, and so at the age of stars like the Sun the increase in peculiar velocity is expected to be about 40 km/s, which is actually somewhat more than deduced from 3D-observations \citep{reid2009}. Therefore, within such a velocity the cosmic ray anisotropy should be small relative to the average interstellar medium around the Solar system.\\
In this paper, we discuss the possible origin of the large-scale structure anisotropies at $\sim \, 10$ TeV with some comments on larger and smaller energies.

\section{Cosmic rays and their Galactic propagation}

Cosmic ray particles are believed to be injected into the ISM by supernova explosions, as soon as their shocks slow down sufficiently to release the population of energetic particles; this may happen for progenitor stars of modest mass as explosions directly into the interstellar medium, or for very massive stars as explosions into their stellar winds. The distinction between these two types of supernova explosions \citep[][and earlier papers]{stanev1993,nath2012} allows to interpret the cosmic ray positron and electron enhancements \citep{biermann2009}, the WMAP haze of high frequency radio emission as well as the $511$ keV annihilation line near the Galactic Center \citep{biermann2010a}, as well as the upturn in the cosmic ray spectra of nuclei \citep{biermann2010b}.
Finally, it allows to understand the KASCADE-Grande spectra of cosmic ray particles at energies beyond $10^{15}$ eV \citep{biermann2012}, as well as allowing to use the cosmic ray particles as injection seeds for the ultra-high energy cosmic rays \citep{gopal2010,biermann2012}.
Then the cosmic ray particles meander, after injection, through the interstellar medium (ISM) subject to the irregularities of the magnetic field. In a general flow not too far from equipartition, the spectrum of the irregularities should be roughly Kolmogorov \citep[e.g.][]{kolmogorov1941,sagdeev1979,goldstein1995}. This has been shown to be consistent with observations \citep{sprangler_gwinn1990}. This has been demonstrated to be consistent with the low level of anisotropies \citep[e.g.][]{biermann1993,blasi_amato2012a,blasi_amato2012b}. 

\subsection{Interaction test with cosmic rays}

However, it is well known, that at the same time, as particles scatter around they also interact, and so the spectrum of the spallation secondaries, like Boron, compared with Carbon, should also reflect the same energy dependence: This is not correct, since the B/C ratio scales as $E^{-0.54}$ \citep{ptuskin1999}; this is inconsistent with a Kolmogorov spectrum. However, it could be consistent with a Kraichnan spectrum $k^{-3/2}$ \citep{kraichnan1965}: However, Kraichnan turbulence implies a lower dimensionality due to a dominant magnetic field, which is not given in the interstellar medium \citep{beck1996}. 
The contradiction can be resolved: The massive stars, that produce most of the CR Carbon before they explode, are Wolf Rayet stars. These stars have a powerful wind and eject most of their zero-age-main-sequence mass into the wind, before they explode \citep{prantzos1984}. This wind forces most of the stellar mass as well as a large amount of ISM material into a thick shell. Therefore at the time of the explosion, typically more than half of the original stellar mass is contained in the wind and its shocked shell, composed of old stellar and ISM material. There, the CRs themselves excite a spectrum of magnetic turbulence \citep{bell1978a,bell1978b}.
\citet{biermann1998} calculated the energy dependence of the escape yielding for the B/C ratio an energy dependence of $E^{-5/9}$ based on the excitation of the irregularities by the cosmic rays themselves \citep{bell1978a,bell1978b} is fully consistent with the observed spectral dependence \citep{ptuskin1999}.
 This suggests that most of the CR interaction happens in these shells. A recent example of observing this cosmic ray interaction directly in $\gamma$-ray data was shown by \citet{berezhko2009}, who argued directly for interaction in a wind-environment. It also implies that there is a minimum interaction, or in other words, a finite minimum path-length of interaction, stemming from the fact that the particles at very high energies do not scatter through the wind-shell, but convect \citep{biermann1993,biermann2001,nath2012}.
In this latter point of a finite residual path length the model is consistent with the results by the Tracer experiment \citep{obermeier2011}, but needs larger statistics data for final confirmation.
There is another test for this argument from a comparison of the electron and proton spectra, which should have %
the same injection slope at TeV energies if arising from the same shock mechanism in the same source class: At energies well above 10 GeV the CR electrons have a spectrum of $E^{-3.26 \pm 0.06}$ up to TeV energies \citep{Wiebel-Sooth1999} %
and clearly have been steepened by synchrotron and inverse Compton losses \citep{kardashev1962}, and so their injection spectrum is $E^{-2.26 \pm 0.06}$. Comparing this with the CR protons \citep[CREAM:][]{yoon2011}, which give a spectrum of about $E^{-2.66 \pm 0.02}$ near TeV energies, the difference to the corrected CR electron spectrum gives the energy dependence of the diffusive escape, $E^{-0.40 \pm 0.06}$, fully consistent with a Kolmogorov law. A spectrum of close to $E^{-8/3}$ matches the prediction for wind-SN CRs (see also \citet{biermann2010b} on the match to the new CREAM data \citep{yoon2011}).
On the other hand, the arguments for the positron fraction rising with energy \citep{biermann2009} support the point of view that wind-SNe are more important for observed CR-electrons at these energies; the predicted ratio of CR-positrons to CR-electrons \citep{biermann2009} has recently been confirmed again by \citet{ackermann2012}.
Therefore, in the following we will use the concept, that for interstellar medium propagation the relevant spectrum of irregularities has a Kolmogorov-type spectrum.

\subsection{Isotropy and anisotropy}

However, before we move on to discuss the anisotropies we should note, that the observation of a near-isotropy is almost more interesting: The isotropy has been used already early to argue that magnetic fields are required to isotropize them \citep{biermann_schlueter1951,biermann1953a,biermann1953b,biermann_davis1958,biermann_davis1960,ginzburg_syrovatskii1964}, and so their strength has been estimated remarkably close to what has been measured since \citep[e.g.][]{beck1996}. But what had never been clear is whether the cosmic ray reference frame is really our frame, the Solar system frame, or any specifically defined local frame, with reference to local stars for instance.
In order to discuss the origin of anisotropies we need to consider what we actually observe: Cosmic ray particles scatter with a mean free path dependent on the spectrum of magnetic irregularities; so in any direction we observe the surface of last effective scattering, and it is critically important to note, that this surfaces itself has a depth equivalent to its distance, so for any specific particle the distance is defined by the length scale over which it forgets its original direction, and comparing different particles that distance is a broad distribution with a width similar to its distance, as indicated by the exponential path-length distribution description \citep{garcia-munoz1987}.
It immediately follows that looking into different directions, by about a radian, these patches of the surface of last effective scattering may well be uncorrelated, with of order 4 patches possible, one of which will be dominant, one will be second, and the rest will produce what looks like noise, unless our statistics become incredibly good. Of course if one recent SN-explosion completely wipes out all traces of previous other SN-explosions then we will be able to detect that in many other traces of recent SN-explosions, possibly even with light-echoes; so individual sources could produce clearly excesses of smaller angular extent. Similarly, magnetic enhancements in specific features, like the heliotail, could produce deficits of smaller angular extent.\\
\begin{figure}[ht]%
 \begin{center}%
 \includegraphics[width=0.9\linewidth]{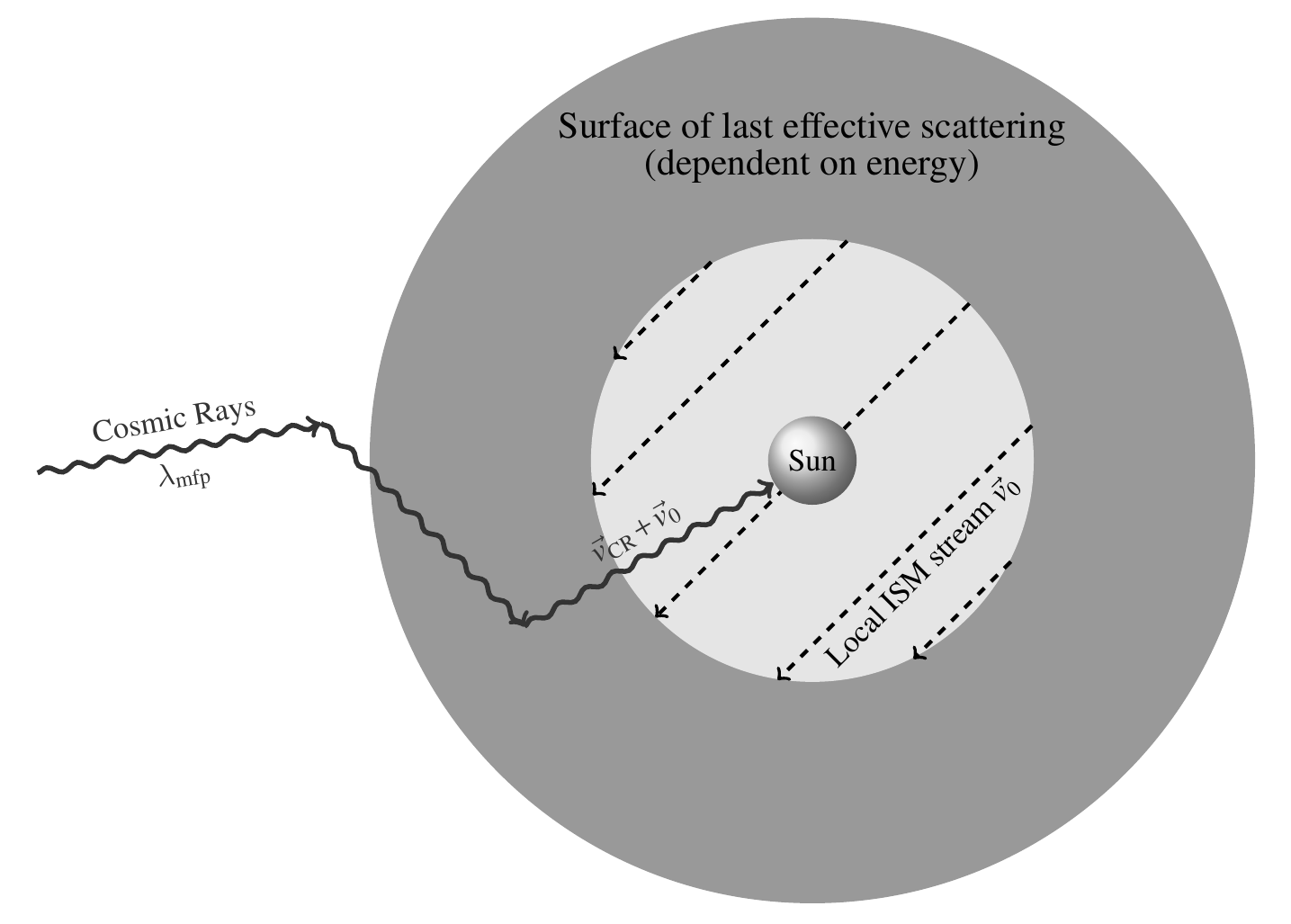}
 \end{center}%
 \label{fig1}%
 \caption{Schematic geometry of the surface of effective last scattering. We note that this surface is itself quite ``thick'' with a range of scattering distances of the same scale as the scattering path itself}
\end{figure}%
Figure 1 shows a schematic picture of the model presented here, emphasizing the role of the surface of last effective scattering.\\
Using a diffusive description of cosmic ray propagation \citet{blasi_amato2012a,blasi_amato2012b} have modelled the possible cosmic ray anisotropies, and have found that the anisotropy could be quite large.\\
Here we propose to take a slightly different route, comparing the influence of both diffusion and velocity fields, always emphasizing that the surface of last effective scattering is ``thick'', as noted; this entails that the higher the energy, the larger the scale, but also the larger the smearing over all lower scale spatial variations. We will focus on the large angle anisotropy near $\sim \, 10$~TeV.\\
The escape from the Galactic disk can be described as a random walk with the step size a function of energy \citep{chandrasekhar1943}. Another related description is with the exponential path-length distribution %
\citep{garcia-munoz1987}. The escape time from the thick Galactic disk \citep{beuermann1985,ferrando1993,brunetti2000} of an approximate half thickness of $d = 1 - 2$~kpc is of order $10^{7}$~yrs, and using the argument above depends on energy as $E^{-1/3}$. Considering a possible heavy composition of Galactic CRs at $10^{18.5}$~eV we will scale to proton energies at $10^{17}$~eV.
Under the cautious assumption that the thickness $H$ is of order three times the mean free path at the maximal energy we obtain a mean free path $\lambda_\mathrm{\rm mfp}$ of order:
\begin{equation}
\lambda_{\rm mfp} = \frac{1}{3} H {\left(\frac{E}{10^{17} {\rm eV}}\right)}^{1/3} \simeq 10^{18.5} {\left(\frac{E}{{\rm GeV}}\right)}^{1/3} {\rm cm}\,.
\end{equation}
Since at that level in vertical direction, about $1 - 2$~kpc, we have a transition to a Galactic magnetic wind \citep{everett2008,everett2010,everett_zweibel2011}, which also scatters particles quite effectively due to the larger scale and the $1/r$-behavior of the magnetic field \citep{parker1958,biermann2012}, this ought to be enough to ensure near isotropy. The scale of order 20 pc is then reached for particle energies of about 10 TeV with an energy dependence as $E^{1/3}$, and a scale of order 70 pc is reached at about 400 TeV.\\
So, let us first consider the various sources of anisotropies, first looking at source distributions, and second at velocity fields. 

\subsubsection{Sources}

The first effect is an overall gradient of cosmic ray particles in a disk galaxy like ours \citep{ginzburg_syrovatskii1964}; however, a disk galaxy is usually a spiral, with some variation across the spiral arms \citep[e.g.][]{beck_hoernes1996}, so that the key radial gradient has to be considered along the spiral arms, so effectively increasing the radial scale of comparison by $\{\cos \theta\}^{-1}$. where $\theta$ is the angle of the spiral arms with respect to a circle. This implies that we expect anisotropy of order one permille, by dividing 20 pc by several times 8 kpc, but symmetric, which is not seen in the data. This probably implies that this effect is washed out out by other influences. Detailed observations at several wavelength in radio and infrared clearly show \citep[e.g.][]{tabatabaei2007}, that the local variations on scales of a few tens of pc and larger dominate the unevenness of the non-thermal radio-emission, and so probably also of the cosmic ray distribution. At much higher energies 
this effect may have been detected \citep[see e.g.][]{teshima2001}, but even there, source regions may dominate what is seen.\\
The second effect is due to the diffusion straight out from the disk into the halo Galactic magnetic wind \citep{everett2008,everett2010,everett_zweibel2011}. However, this effect is of second order due to the symmetry, and depends critically on where the mid-plane is for this flow and diffusion, which we do not know. Assuming for lack of a good number for the distance from Earth to mid-plane of the flow-field about 40 pc, and for the vertical scale of the flow again 1.5 kpc, suggests an asymmetry of order $6 \times 10^{-4}$, no longer visible at any length even approaching 40 pc. This again is clearly not dominant in what we observe.\\
Finally, the obvious most recent source, an idea which has been explored very many times \citep[e.g.][]{voelk1988a,teshima2001,erlykin_wolfendale2006,yueksel2009}. One difficulty with such an idea is that there is no clear spectral signature of any recent source of cosmic rays; on the other hand, such a source must exist, just statistically. The question is, what is its signature? The supernova rate in our Galaxy has been estimated to be of order 1 per 30 years, with an uncertainty towards longer times of a factor of 3 \citep[see e.g.][]{biermann-cas1993,biermann1995}; averaging this uncertainty blithely suggests that one supernova in 50 years, to within a factor of 2, might be a good initial estimate. The probability to then have a supernova explode within a certain radial distance over a certain time, at our distance from the Galactic Center, can then be estimated. Since the supernova rate is reduced at our distance from the Galactic Center with respect to the average can be crudely 
estimated to be of order 3, so that in our part of the Galaxy, the supernova rate may be about one in 150 years. It follows that within a distance of $r \, = \, R_{1.3} \, 20$~pc from us the time scale between supernovae is of order $10^{7.6} \, R_{1.3}^{-2}$~yrs.\\
This, then, allows to derive the diffusion coefficient derived above as $10^{29.8} \, \mathrm{cm}^{2} \mathrm{s}^{-1}$ at 10 TeV; this is slightly larger than the estimate given by \citet{blasi_amato2012a,blasi_amato2012b}, who use a different concept of cosmic ray propagation, but our number is close to older estimates. The uncertainties are dominated by systematics, since different lines of reasoning have been used to derive the coefficient as well as its energy dependence.\\
This implies that over a distance of 20 pc the diffusive time-scale is of order $10^{2.4} \, R_{1.3}^{2}$ yrs. Therefore the dispersion of any original cloud of CR-particles is extremely effective, with the typical time scale between supernovae taken from above we obtain a radial range of many kpc numerically, and a dilution relative to the average cosmic ray energy density of about $10^{-5.4}$, so the typical anisotropy expected from supernovae should be quite small. This predicted anisotropy scales with the reference distance as $R_{1.3}^{+3}$, so for a mean free path at 10 TeV larger by a factor of two, well within the uncertainties, we obtain a predicted cosmic ray anisotropy of $10^{-4.5}$, getting close to numbers observed. This clearly implies that a recent supernova explosion cannot be immediately discounted as a contributing origin for cosmic ray anisotropies, confirming \citep[e.g.][and others]{voelk1988a,teshima2001,erlykin_wolfendale2006,yueksel2009}.

\subsubsection{Residual flow fields}

However, when we consider the flow field produced by supernova explosions, then the numbers change, since the flow is just mixed slowly after decaying to subsonic and sub-Alfv{\'e}nic velocities.\\
The simplest second reason for cosmic ray particles to show anisotropy is a general movement of our frame of reference with respect to that of the cosmic ray system \citep{compton1935}.\\
Data and magneto-hydrodynamic simulations suggest that the ISM is mainly driven by supernova explosions, phase transitions, cloud and star formation, radiation from young stars, instabilities and energy transport by magnetic fields and CR particles, and shear and outflow from the Galactic disk \citep{cox1972,parker1966,mckee1977,breitschwerdt1991,lee2003,hanasz2004,everett2008,everett2010,everett_zweibel2011}.
 The hottest normal phase of the ISM is at about a density of $3 \times 10^{-3} \; \mathrm{cm}^{-3}$ and a temperature of about $4 \times 10^{6}$~K \citep{snowden1997,hagihara2011}, confirming earlier expectations \citep{lagage1983}.
 After a supernova explodes into the ISM, the expansion of the shock racing through the ISM is fast at first, accumulating evermore material from the ISM, and then slows down \citep{sedov1958,cox1972}, until the flow becomes subsonic and sub-Alfv{\'e}nic; thereupon the flow coasts along, and basically gets slowly disorganized by mixing and encountering clouds. \citet{gaensler2011} confirm that the typical sonic Mach-number is low, of order 2 or less. At the density and temperature of this most tenuous phase the signal speeds will thus be around 160 km/s, and so the velocity fields of the late evolution of supernova remnants (SNRs) will run at or a bit below these velocities, but later decay rather slowly. The average density across all media in the 200 pc thick layer of the ISM is of order $1 \; \mathrm{cm}^{-3}$ \citep{cox2005}. Using this average density we obtain a velocity scale of order 100 km/s at a scale of about 30 pc, using the simple expressions of \citet{cox1972}.
 The associated time scale is of order $2 \times 10^{5}$~yrs. Thus, the SNR is no longer supersonic or super-Alfv{\'e}nic in the hot medium. Observational data on pulsar activity and their associated SNRs support the scale of 30 pc \citep{braun1989}, and a time scale for the powering of $2\times 10^{4}$~yrs, less than $2 \times 10^{5}$~yrs, so consistent with the numbers suggested here for the hot interstellar gas.
 Similarly, the confinement time for CR particles in decaying SNRs has been estimated also to be relatively short, of order $10^{4}$ yrs \citep{berezhko2004}, in agreement with the arguments by \citet{braun1989}. Cosmic rays will travel most easily in that phase of the ISM, where the Alfv{\'e}n velocity is the highest, due to limiting the 
streaming instability at the Alfv{\'e}n velocity. Therefore, we might expect the CR particles that we observe to come most effectively through this phase. However, as the observations show \citep{appenzeller1974} the magnetic fields permeate the clouds, and so it is more suitable to use the average density of $1 \; \mathrm{cm}^{-3}$. So the old velocity field of SNRs having gone subsonic and sub-Alfv{\'e}nic in the highest temperature medium corresponds to velocities of about 100 km/s and to length scales of about 30 pc.\\
Considering the scattering of CRs arriving at Earth, at such a distance from us when they first point in our general direction, that surface can be called {\it the surface of last effective scattering}, one mean free path away. If that region has a general flow field, it will imprint an anisotropy upon the CRs coming to Earth. The observed CR anisotropy corresponds maximally to about $10^{-3}$; comparing with the anisotropy of $3 \times 10^{-4}$ attributed to the motion of the Earth around the Sun of 30 km/s, this implies a velocity amplitude of about 100 km/s. This matches the velocities of %
old SNRs after they become sub-sonic and sub-Alfv{\'e}nic, using the Compton-Getting effect \citep{compton1935}, which connects velocity field $v$, CR spectrum $E^{-p} \, \rm{d} E$ and CR anisotropy $\Delta I / I_{\rm av}$:
\begin{equation}
\frac{\Delta I}{I_{\rm av}} \; = \; \frac{v}{c} \, (p + 2) \, \cos \theta.
\end{equation}
It has to be noted, that this effect does not require that we sit in the center of a SNR: The effect is there as soon as we have coherent motion at the surface of last scattering over about one radian in one direction laterally, and in depth by about the distance itself, so $\Delta r / r \sim 1$; consistent with this the data suggest an effect which varies across the sky. Massive stars that later produce supernovae form in the cold disk which is only of order 200 pc full width 
\citep{cox1974,cox2005}. Using the time scale from above of about 200,000 yrs yields a length scale about a factor of 2 larger, with the energetics adopted by \citet{cox1972}.
The irregular flow field will be dominated by a single most recent supernova up to this scale combined with several supernovae that are somewhat older. However, using this time scale we need to go back to our earlier question and ask, what the source anisotropy could be at such a time scale, and the answer is $10^{-3.6}$, so in agreement with \citet{blasi_amato2012a,blasi_amato2012b} not completely negligible compared to the influence of the flow itself. And also, just like the flow the directionality is not isotropic, the only key difference is, that the source contribution can only be positive, but the modified Compton-Getting effect can be either positive or negative; the observations suggest at 400 TeV that the major effect is a deficit, consistent with a modified Compton-Getting effect.\\
To summarize in yet another way, for any scale $L < H$, where $H$ is the cosmic ray scale-height \citep{biermann2001}, we have the inequality

\begin{equation}
\frac{L^2}{\kappa} \, < \, \frac{L}{V_\mathrm{signal}}
\end{equation}

where $V_\mathrm{signal}$ is either the speed of sound $c_s$, the Alfv{\'e}n velocity $V_A$, or some combination thereof (the fast magneto-sonic speed). On the left side this is the cosmic ray diffusion time scale over a length $L$, and on the right hand side this is the convection time scale over the same length scale. At $L \, = \, H$ the two time scales become equal, and convection takes over, and the loss from the disk is then driving a convective Galactic wind \citep{breitschwerdt1991,everett2008,everett2010,everett_zweibel2011,biermann2010a}. However, it then also follows that convective flow is slower for all $L \, < \, H$, so lives longer, and can dominate cosmic ray scattering, and so is more viable to let us understand the observed anisotropy.

\subsubsection{Magnetic field isotropy?}

There are several other effects which need to be considered in the propagation of CRs.
First of all, the spectrum of irregularities may be well approximated by an isotropic Kolmogorov spectrum, but since the observed magnetic field distribution is never completely irregular \citep{beck1996} there can always be effects from this underlying anisotropy. Furthermore, even the irregularities themselves are never perfectly isotropic either \citep[e.g.,][]{malkov2010,lazarian2010,desiati_lazarian2011}, and so such an assumption has to be taken with great caution, even though it may appear that it works relatively well. Finally, and perhaps most importantly, the history of recent supernova explosions in the Solar neighborhood will give an imprint of irregularity just from the source distribution; however, since the time to escape is much larger than the time to replenish the cosmic ray population this should perhaps not be dominant until one gets to really high energies.\\
The sky maps in radio rotation measure \citep{oppermann2011}, in radio emission \citep{berkhuijsen1971} and other wavelengths usually integrate over much larger distances, that it is difficult to be sure of a correlation with the detected anisotropies. There may be a correlation of some radio features like {\it Spur 185-} \citep{berkhuijsen1971} with features in Tibet, Milagro, SuperK and IceCube data;
there may be a correlation with {\it Loop I} in IceCube data, and {\it Loop IV} in Tibet data; and finally, there may be a correlation with {\it Spur 195+} in Milagro data. The deficit in the IceCube flux at 20 TeV is remarkably close to the center of {\it Loop I}. It needs however a careful analysis of the cosmic ray data and a detailed modeling of the theory in order to investigate the causality of the signatures; today's radio data are much better \citep[e.g.][]{oppermann2011,van_eck2011,pshirkov2011}, but what we need is yet more significant cosmic ray data.\\
After subtracting the large angle features it is possible to identify small angular scale features \citep{abdo2008}, which may related to Solar wind effects such as the heliotail \citep{nagashima1998,drury2008,lazarian2010,desiati_lazarian2011}, or nearby stars like Aldebaran \citep{van-Leeuwen1998}, or even nearly extinct pulsar tails (\citet[e.g.][]{romanova2005}). IBEX data \citep{mccomas2009,mccomas2011} show that the Solar wind interacts with the ISM, and emits a steady stream of low energy neutral atoms, producing a ribbon in the sky, completely unanticipated by theory and earlier observations. Here, we do not address these small scale features but focus on the explanation of the large-scale anisotropy.\\
It has been argued that we sit in a local bubble, the walls of which are made up of various old SNR shells. However, a detailed study \citep{mebold1998} suggests that this apparent bubble is not a coherent figure, but rather a motley assembly of filaments, clouds and shells, assembled into a bubble only in perception. These features represent the local history of star formation, HII regions, the effect of other stellar activity, and old violent supernovae.

\subsection{Balance of effects on cosmic ray anisotropy}

We conclude that old supernova remnants leave traces of their flow-field for much longer times, than traces of their individual cosmic ray contribution. Therefore, the population of cosmic ray particles in its scattering is more strongly influenced by these flow fields, than by the inhomogeneity of the sources in space and time, unless the source is unusually young. The anisotropy given by the flow fields is usually not dipole-like, since looking into different directions in the sky will see the thick surface of last effective scattering usually in different old supernova remnant flow-fields.\\
It follows that there is an optimum particle energy to detect anisotropies, and at any larger energies we begin to smooth over several irregular flow patches (which clearly shows that the diffusion approximation is no longer adequate in describing anisotropies), so that the anisotropies become smaller. Also, going to significant larger scales we begin to see different flow patches, and so the directionality would become uncorrelated at larger energies and thus larger scales as seen by inverting the scale energy connection: $E \sim \lambda_{\rm mfp}^{3}$; this is derived by inverting the Kolmogorov spectrum with the exponent 1/3.
The IceCube data are consistent with both these effects, as the anisotropy is weaker at higher energies, and also uncorrelated in direction. Finally we note that of course the Sun may have a peculiar velocity with respect to its environment: This has been shown by VLBI observations to be a small effect, $< \, 20 \;\mathrm{km}/\mathrm s$ \citep{reid2009}. However, we need to emphasize that the uncertainties are large in such simple arguments.
Thus, we need to match particle energy and expected SNR scale: The old SNRs of about 30 pc match the mean free path for about 10 TeV. This should then correspond roughly to the maximum of any anisotropy; at larger spatial scales and thus larger energies the anisotropy is smeared out across several old SNRs; at lower energies we reach portions of an old SNR, not giving the full amplitude. Another important aspect is that in this model there is no expectation of symmetry, since in one direction and the directly opposite direction the surface of last effective scattering will touch different old supernova remnants.\\
However, how does this compare to the elaborate calculations of \citet{blasi_amato2012a,blasi_amato2012b}? These authors use the diffusion approximation, which of course begins to fail when the scale of the inhomogeneities becomes similar to the scale of the mean free path of scattering itself, a case which we argue we have here. So, using the diffusion approximation increases the apparent anisotropy when the scale of old supernova remnants is reached, while treating the scattering directly it becomes obvious that then the ``thickness'' of the surface of last effective scattering itself begins to smear out anisotropies and so decreases it.\\
Another aspect is that counting nearby sources of cosmic rays, when old supernova remnants not just produce a flow field, but of course also slowly disperse the cosmic rays as a function of time, produces a gradient in cosmic rays, which would also give an anisotropy: However, the data suggest that the aspect is a significant deficit, but not a significant excess \citep{abbasi2011b}, and this is more easily explained as a Compton-Getting effect \citeyearpar{compton1935}, which can have either sign, but does not have to be symmetric, as noted above.\\
A key assumption in this model is the use of a Kolmogorov spectrum for the irregularities across the entire energy range of Galactic CRs, from $10^{9}$ eV to $10^{18}$ eV, a point on which we agree with \citet{blasi_amato2012a,blasi_amato2012b}. Only this assumption allows the match in length scales.\\
The amplitude at around 10 TeV would then be of order three times what is seen for the motion of the Earth around the Sun, so about $10^{-3}$ at most, and would correspond to angular scales of about 60 degrees. At larger energies the anisotropies begin to get smeared out. This is consistent with what is observed.

\section{Conclusions}

In summary, we note as others have done before, that a variety of effects may contribute to an observed anisotropy. The anisotropy suggested here, due to the Compton-Getting effect in the magnetized flow field of old slowly disappearing SNRs would give rather large angle scale fluctuations, since the mean free path is valid in all spatial directions, and so angular scales, as noted above should be of order 60 degrees. The effect is a sum of uncorrelated monopole components. That feature distinguishes this proposal from most other possibilities, which can easily produce small scales. The key here was to assume that a Kolmogorov spectrum gives the length scale for a particle energy, which is quite different from other possible scattering descriptions.\\
To the measure that the old SNRs argued here contribute to the main flow field influencing the surface of last effective scattering still have some of their own contribution to CRs, one might expect that contribution to be slightly flatter in its CR spectrum, at those locations in the sky, where the anisotropy is most visible.\\
The effect suggested here predicts that at energies apart by more than an order of magnitude the anisotropies should become uncorrelated, unless we sit in a very large-scale substantial flow-field.
Averaging over several typical old SNRs scales the anisotropies should decrease with energy. Finally, to first order the anisotropies near about 10 TeV should have angular scales of order 60 degrees.\\[5mm]
Discussions with E.\ Berkhuijsen, R.\ Engel, H.\ Falcke, P.P.\ Kronberg, A.\ Lazarian, V.\ Ptuskin, Ch.\ Spiering, and Ch.\ Wiebusch are gratefully acknowledged by PLB, with R.\ Abbasi, J.\ Black, P.\ Desiati, M.\ Santander, S.\ Toscano, St.\ Westerhoff, and M.\ Zhang by JKB, with J.R.\ Jokipii and M.-A.\ Malkov by ESS. JKB also wishes to express her appreciation to the discussions with the IceCube Collaboration; Support for JKB and MM comes from the DFG grant BE 3714/5-1 as a part of FOR1048 (``Instabilities, Turbulence and Transport in Cosmic Magnetic Fields'') and from the Research Department of Plasmas with Complex Interactions (Bochum). Support for ESS comes from NASA grant NNX09AC14G.


\begin{thebibliography}{99}
\providecommand{\natexlab}[1]{#1}
\providecommand{\url}[1]{\texttt{#1}}
\expandafter\ifx\csname urlstyle\endcsname\relax
 \providecommand{\doi}[1]{doi: #1}\else
 \providecommand{\doi}{doi: \begingroup \urlstyle{rm}\Url}\fi
\bibitem[{Abbasi} et~al.(2010)]{abbasi2010}
R.~{Abbasi} et~al.
\newblock \emph{\apjl}, 718:\penalty0 L194, 2010.

\bibitem[{Abbasi} et~al.(2011)]{abbasi2011a}
R.~{Abbasi} et~al.
\newblock \emph{\prd}, 83\penalty0 (1):\penalty0 012001, 2011.

\bibitem[{Abbasi} et~al.(2012)]{abbasi2011b}
R.~{Abbasi} et~al.
\newblock \emph{\apj}, 746:\penalty0 33, 2012.

\bibitem[{Abdo} et~al.(2008)]{abdo2008}
A.~A. {Abdo} et~al.
\newblock \emph{\prl}, 101\penalty0 (22):\penalty0 221101, 2008.

\bibitem[{Abdo} et~al.(2009)]{abdo2009}
A.~A. {Abdo} et~al.
\newblock \emph{\apj}, 698:\penalty0 2121, 2009.

\bibitem[{Ackermann} et~al.(2012)]{ackermann2012}
M.~{Ackermann} et~al.
\newblock \emph{\prl}, 108:\penalty0 011103, 2012.

\bibitem[{Amenomori} et~al.(2006)]{amenomori2006}
M.~{Amenomori} et~al.
\newblock \emph{Science}, 314:\penalty0 439, 2006.

\bibitem[{Appenzeller}(1974)]{appenzeller1974}
I.~{Appenzeller}.
\newblock \emph{\aap}, 36:\penalty0 99, 1974.

\bibitem[{Beck} et~al.(1996)]{beck1996}
R.~{Beck} et~al.
\newblock \emph{\araa}, 34:\penalty0 155, 1996.

\bibitem[{Beck} and Hoernes(1996)]{beck_hoernes1996}
R.~{Beck} and P.\ Hoernes
\newblock \emph{\nat}, 379:\penalty0 47, 1996.

\bibitem[{Bell}(1978{\natexlab{a}})]{bell1978a}
A.~R. {Bell}.
\newblock \emph{\mnras}, 182:\penalty0 147, 1978{\natexlab{a}}.

\bibitem[{Bell}(1978{\natexlab{b}})]{bell1978b}
A.~R. {Bell}.
\newblock \emph{\mnras}, 182:\penalty0 443, 1978{\natexlab{b}}.

\bibitem[{Berezhko} and {V{\"o}lk}(2004)]{berezhko2004}
E.~G. {Berezhko} and H.~J. {V{\"o}lk}.
\newblock \emph{\apj}, 611:\penalty0 12, 2004.

\bibitem[{Berezhko} et al.(2004)]{berezhko2009}
E.~G. {Berezhko}, G.\ P\"uhlhofer and H.~J. {V{\"o}lk}.
\newblock \emph{\aap}, 505:\penalty0 641, 2009.

\bibitem[{Berkhuijsen} et~al.(1971)]{berkhuijsen1971}
E.~M.\ {Berkhuijsen} et~al.
\newblock \emph{\aap}, 14:\penalty0 252, 1971.

\bibitem[{Beuermann} et~al.(1985)]{beuermann1985}
K.~{Beuermann} et~al.
\newblock \emph{\aap}, 153:\penalty0 17, 1985.

\bibitem[L.~{Biermann} and {Schl\"uter}(1951)]{biermann_schlueter1951}
L. {Biermann} and A.~{Schl\"uter}.
\newblock \emph{Phys.\ Rev.}, 82:\penalty0 863, 1951.

\bibitem[L.~{Biermann} and {Davis}(1958)]{biermann_davis1958}
L. {Biermann} and L.~{Davis}.
\newblock \emph{\ZfNat}, 13a:\penalty0 909, 1958.

\bibitem[L.~{Biermann} and {Davis}(1960)]{biermann_davis1960}
L. {Biermann} and L.~{Davis}.
\newblock \emph{\ZfA}, 51:\penalty0 19, 1960.

\bibitem[L.~{Biermann}(1953a)]{biermann1953a}
L. {Biermann}.
\newblock \emph{\ARNSci}, 2:\penalty0 335, 1953.

\bibitem[L.~{Biermann}(1953b)]{biermann1953b}
L. {Biermann}.
\newblock in \emph{Vortr\"age \"uber Kosmische Strahlung}, Springer, Berlin, p.\ 9, 2nd edition, 1953.

\bibitem[{Biermann} and {Cassinelli}(1993)]{biermann-cas1993}
P.~L. {Biermann} and J.~P. {Cassinelli}
\newblock \emph{\aap}, 277:\penalty0 691, 1993.

\bibitem[{Biermann}(1993)]{biermann1993}
P.~L. {Biermann}.
\newblock \emph{\aap}, 271:\penalty0 649, 1993.

\bibitem[Biermann(1998)]{biermann1998}
P.~L. Biermann.
\newblock Cosmic ray interactions in the Galaxy, invited
 lecture at the Nuclear Astrophysics meeting at Hirschegg, in Proc., GSI,
 Darmstadt, p.\ 211, 1998

\bibitem[{Biermann} and {de Souza}(2012)]{biermann2012}
P.~L. {Biermann} and V.~{de Souza}.
\newblock \emph{\apj}, 746:\penalty0 72, 2012.

\bibitem[{Biermann} et~al.(1995)]{biermann1995}
P.~L. {Biermann}, T.\ K.\ Gaisser and T.\ Stanev.
\newblock \emph{\prd} 51:\penalty0 3450, 1995

\bibitem[{Biermann} et~al.(2001)]{biermann2001}
P.~L. {Biermann} et~al.
\newblock \emph{\aap}, 369:\penalty0 269, 2001.

\bibitem[{Biermann} et~al.(2009)]{biermann2009}
P.~L. {Biermann} et~al.
\newblock \emph{\prl}, 103\penalty0 (6):\penalty0 061101, 2009.

\bibitem[{Biermann} et~al.(2010a)]{biermann2010a}
P.~L. {Biermann} et~al.
\newblock \emph{\apjl}, 710:\penalty0 L53, 2010.

\bibitem[{Biermann} et~al.(2010b)]{biermann2010b}
P.~L. {Biermann} et~al.
\newblock \emph{\apj}, 725:\penalty0 184, 2010.

\bibitem[{Blasi} and {Amato}(2012a)]{blasi_amato2012a} P.\ {Blasi} and E.\ Amato.
\newblock \emph{\jcap} 01:\penalty0 010, 2012a.

\bibitem[{Blasi} and {Amato}(2012b)]{blasi_amato2012b} P.\ {Blasi} and E.\ Amato.
\newblock \emph{\jcap} 01:\penalty0 011, 2012b.

\bibitem[{Braun} et~al.(1989)]{braun1989}
R.~{Braun} et~al.
\newblock \emph{\apj}, 340:\penalty0 355, 1989.

\bibitem[{Breitschwerdt} et~al.(1991)]{breitschwerdt1991}
D.~{Breitschwerdt} et~al.
\newblock \emph{\aap}, 245:\penalty0 79, 1991.

\bibitem[{Brunetti} and {Codino}(2000)]{brunetti2000}
M.~T. {Brunetti} and A.~{Codino}.
\newblock \emph{\apj}, 528:\penalty0 789, 2000.

\bibitem[{Chandrasekhar}(1943)]{chandrasekhar1943}
S.~{Chandrasekhar}.
\newblock \emph{\rmp}, 15:\penalty0 1, 1943.

\bibitem[{Compton} and {Getting}(1935)]{compton1935}
A.~H. {Compton} and I.~A. {Getting}.
\newblock \emph{\pr}, 47:\penalty0 817, 1935.

\bibitem[{Cox}(1972)]{cox1972}
D.~P. {Cox}.
\newblock \emph{\apj}, 178:\penalty0 159, 1972.

\bibitem[{Cox}(2005)]{cox2005}
D.~P. {Cox}.
\newblock \emph{\araa}, 43:\penalty0 337, 2005.

\bibitem[{Cox} and {Smith}(1974)]{cox1974}
D.~P. {Cox} and B.~W. {Smith}.
\newblock \emph{\apjl}, 189:\penalty0 L105, 1974.

\bibitem[{Desiati} and {Lazarian}(2011)]{desiati_lazarian2011}
P.\ {Desiati} and A.\ {Lazarian}.
\newblock \emph{arXiv:1111.3075}, 2011.

\bibitem[{Drury} and {Aharonian}(2008)]{drury2008}
L.~O'C. {Drury} and F.~A. {Aharonian}.
\newblock \emph{Astroparticle Physics}, 29:\penalty0 420, 2008.

\bibitem[{van Eck} et al.(2011)]{van_eck2011}
C.~L.~{van Eck} et al.
\newblock \emph{\apj}, 728:\penalty0 97, 2011.

\bibitem[{Erlykin} and Wolfendale(2006)]{erlykin_wolfendale2006}
A.\ D.\ {Erlykin} and A.\ W.\ Wolfendale
\newblock \emph{\app}, 25:\penalty0 183, 2006.

\bibitem[{Everett} et~al.(2008)]{everett2008}
J.~E. {Everett} et~al.
\newblock \emph{\apj}, 674:\penalty0 258, 2008.

\bibitem[{Everett} et~al.(2010)]{everett2010}
J.~E. {Everett} et~al.
\newblock \emph{\apj}, 711:\penalty0 13, 2010.


\bibitem[{Everett} and {Zweibel}(2011)]{everett_zweibel2011}
J.~E.\ {Everett} and E.\ G.\ {Zweibel}.
\newblock \emph{\apj}, 739:\penalty0 60, 2011.

\bibitem[Ferrando(1993)]{ferrando1993}
P.~Ferrando.
\newblock In \emph{23rd ICRC (Calgary)}, page 279, 1993.

\bibitem[{Gaensler} et~al.(2011)]{gaensler2011}
B.~M. {Gaensler} et~al.
\newblock \emph{\nat}, 478:\penalty0 214, 2011.

\bibitem[{Garcia-Mu\~{n}oz} et~al.(1987)]{garcia-munoz1987}
M.~{Garcia-Mu\~{n}oz} et~al.
\newblock \emph{\apjs}, 64:\penalty0 269, 1987.

\bibitem[{Ginzburg} and {Syrovatskii} (1964)]{ginzburg_syrovatskii1964}
V.\ L.\ {Ginzburg} and S.\ I.\ {Syrovatskii}.
\newblock In \emph{The origin of cosmic rays}, Pergamon, Oxford 1964; origin Russian 1963.

\bibitem[{Goldstein} et~al.(1995)]{goldstein1995}
M.~L. {Goldstein} et~al.
\newblock \emph{\araa}, 33:\penalty0 283, 1995.

\bibitem[{Gopal-Krishna} et~al.(2010)]{gopal2010}
{Gopal-Krishna} et~al.
\newblock \emph{\apjl}, 720:\penalty0 L155, 2010.

\bibitem[{Guillian} et~al.(2007)]{guillian2007}
G.~{Guillian} et~al.
\newblock \emph{\prd}, 75\penalty0 (6):\penalty0 062003, 2007.

\bibitem[{Hagihara} et~al.(2011)]{hagihara2011}
T.~{Hagihara} et~al.
\newblock \emph{\pasj}, 63:\penalty0 889, 2011.

\bibitem[{Hanasz} et~al.(2004)]{hanasz2004}
M.~{Hanasz} et~al.
\newblock \emph{\apjl}, 605:\penalty0 L33, 2004.

\bibitem[{Hanasz} et~al.(2009)]{hanasz2009}
M.~{Hanasz} et~al.
\newblock \emph{\aap}, 498:\penalty0 335, 2009.
%

\bibitem[{Julian}(1967)]{julian1967}
W.~H.\ {Julian}.
\newblock \emph{\apj}, 148:\penalty0 175, 1967.

\bibitem[{Kardashev}(1962)]{kardashev1962}
N.~S. {Kardashev}.
\newblock \emph{\azh}, 39:\penalty0 393, 1962.

\bibitem[{Kolmogorov}(1941)]{kolmogorov1941}
A.~{Kolmogorov}.
\newblock \emph{Akad.\ Nauk SSSR Dokl.}, 30:\penalty0 301, 1941.

\bibitem[{Kraichnan}(1965)]{kraichnan1965}
R.~H.\ {Kraichnan}.
\newblock \emph{Phys.\ Fl.}, 8:\penalty0 1385, 1965.

\bibitem[{Lagage} and {Cesarsky}(1983)]{lagage1983}
P.~O. {Lagage} and C.~J. {Cesarsky}.
\newblock \emph{\aap}, 125:\penalty0 249, 1983.

\bibitem[{Lazarian} and {Desiati}(2010)]{lazarian2010}
A.~{Lazarian} and P.~{Desiati}.
\newblock \emph{\apj}, 722:\penalty0 188, 2010.

\bibitem[{Lee} et~al.(2003)]{lee2003}
H.~{Lee} et~al.
\newblock \emph{\apj}, 594:\penalty0 627, 2003.

\bibitem[{van Leeuwen} and {Evans}(1998)]{van-Leeuwen1998}
F.~{van Leeuwen} and D.~W. {Evans}.
\newblock \emph{\aaps}, 130:\penalty0 157, 1998.

\bibitem[{Malkov} et~al.(2010)]{malkov2010}
M.~A. {Malkov} et~al.
\newblock \emph{\apj}, 721:\penalty0 750, 2010.

\bibitem[{McComas} et~al.(2009)]{mccomas2009}
D.~J. {McComas} et~al.
\newblock \emph{Science}, 326:\penalty0 959, 2009.

\bibitem[{McComas} et~al.(2011)]{mccomas2011}
D.~J. {McComas} et~al.
\newblock \emph{\grl}, 38:\penalty0 L18101, 2011.

\bibitem[{McKee} and {Ostriker}(1977)]{mckee1977}
C.~F. {McKee} and J.~P. {Ostriker}.
\newblock \emph{\apj}, 218:\penalty0 148, 1977.

\bibitem[{Mebold} et~al.(1998)]{mebold1998}
U.~{Mebold} et~al.
\newblock In {D.~Breitschwerdt, M.~J.~Freyberg, \& J.~Truemper}, editor,
 \emph{IAU Colloq. 166}, volume 506, page 199, 1998.

\bibitem[{Nagashima} et~al.(1998)]{nagashima1998}
K.~{Nagashima} et~al.
\newblock \emph{\jgr}, 103:\penalty0 17429, 1998.

\bibitem[{Nath} et~al.(2012)]{nath2012}
B.~B. {Nath}, N. {Gupta} and P.~L. {Biermann}
\newblock \emph{arXiv:1204.4239}, 2012.

\bibitem[{Obermeier}(2011)]{obermeier2011}
A.~{Obermeier}.
\newblock PhD thesis, Radboud University Nijmegen, 2011.

\bibitem[{Oppermann} et~al.(2011)]{oppermann2011}
N.~{Oppermann} et~al.
\newblock \emph{arXiv:1111.6186}, 2011.

\bibitem[{Parker}(1958)]{parker1958}
E.~N. {Parker}.
\newblock \emph{\apj}, 128:\penalty0 664, 1958.

\bibitem[{Parker}(1966)]{parker1966}
E.~N. {Parker}.
\newblock \emph{\apj}, 145:\penalty0 811, 1966.

\bibitem[{Prantzos}(1984)]{prantzos1984}
N.~{Prantzos}.
\newblock \emph{Adv.\ in Space Res.}, 4:\penalty0 109, 1984.

\bibitem[{Pshirkov}(2011)]{pshirkov2011}
M.~S.\ {Pshirkov}.
\newblock \emph{\apj}, 738, 192, 2011.

\bibitem[{Ptuskin}(1999)]{ptuskin1999}
V.~{Ptuskin}.
\newblock In \emph{ICRC}, volume~4, page 291, 1999.

\bibitem[{Reid} et~al.(2009)]{reid2009}
M.~J. {Reid} et~al.
\newblock \emph{\apj}, 700:\penalty0 137, 2009.

\bibitem[{Romanova} et~al.(2005)]{romanova2005}
M.~M. {Romanova} et~al.
\newblock \emph{\apj}, 630:\penalty0 1020, 2005.

\bibitem[{Sagdeev}(1979)]{sagdeev1979}
R.~Z. {Sagdeev}.
\newblock \emph{\rmp}, 51:\penalty0 1, 1979.

\bibitem[{Schl\"uter} and {Biermann} (1950)]{schlueter_biermann1950}
A.\ Schl\"uter and L.\ {Biermann}.
\newblock \emph{\ZNat}, 5a:\penalty0 237, 1950.

\bibitem[{Sedov}(1958)]{sedov1958}
L.~I. {Sedov}.
\newblock \emph{\rmp}, 30:\penalty0 1077, 1958.

\bibitem[{Snowden} et~al.(1997)]{snowden1997}
S.~L. {Snowden} et~al.
\newblock \emph{\apj}, 485:\penalty0 125, 1997.

\bibitem[{Sprangler} and {Gwinn}(1990)]{sprangler_gwinn1990} S.\ R.\ Spangler and C.\ R.\ Gwinn.
\newblock \emph{\apjl}, 353:\penalty0 L29, 1990.

\bibitem[{Stanev} et al.(1993)]{stanev1993} T.\ Stanev, P.~L.~Biermann and T.\ K.\ Gaisser.
\newblock \emph{\aap}, 274:\penalty0 902, 1993.

\bibitem[{Tabatabaei} et al.(2007)]{tabatabaei2007} F.\ S.\ Tabatabaei et al.
\newblock \emph{\aap}, 475:\penalty0 133, 2007.

\bibitem[{Teshima} et al.(2001)]{teshima2001} M.\ Teshima et al.
\newblock in: \emph{Proc.\ 27th ICRC}, p.337, 2001.

\bibitem[{V{\"o}lk} et~al.(1988)]{voelk1988a}
H.~J. {V{\"o}lk} et~al.
\newblock \emph{\aap}, 198:\penalty0 274, 1988.

\bibitem[{Wielen}(1975)]{wielen1975}
R.\ {Wielen}.
\newblock In \emph{La dynamique des galaxies spirales}, Ed.\ L.\ Weliachew, Editions du CNRS, Paris, p.\ 357, 1975.

\bibitem[{Wiebel-Sooth} and {Biermann}(1999)]{Wiebel-Sooth1999}
B.~{Wiebel-Sooth} and P.L. {Biermann}.
\newblock \emph{Cosmic Rays}, page~37.
\newblock Springer Publ. Comp., 1999.

\bibitem[{Yoon} et~al.(2011)]{yoon2011}
Y.~S. {Yoon} et~al.
\newblock \emph{\apj}, 728:\penalty0 122, 2011.

\bibitem[{Y\"uksel} et~al.(2009)]{yueksel2009}
H.\ {Y\"uksel}, M.\ D.\ Kistler and T.\ Stanev
\newblock \emph{\prl}, 103:\penalty0 051101, 2009.
\end{thebibliography}
\end{document}